\documentstyle[twoside,epsfig]{hsproc}
\def\runauthor{R. Lednick\'y}
\def\shorttitle{Multi-boson effects and space-time asymmetries}
\begin{document}
\title{Searching for multi-boson effects and space-time asymmetries 
in particle production}
\author{R. Lednick\'y}{Institute of Physics, Na Slovance 2, 18221 Prague 8,
Czech Republic}
%
\abstract{The influence of the multi-boson effects on pion multiplicities,
single-pion spectra and two-pion correlation functions is discussed
in terms of an analytically solvable model.
It is argued that spectacular multi-boson effects are likely to be observed
only in the rare events containing sufficiently high density
fluctuations.
The possibilities of unlike particle
correlations for a study of the space-time asymmetries in particle
production, including the sequence of particle emission, 
are demonstrated.}
\section{Introduction}
The correlations of particles at small relative velocities
are widely used to study space-time characteristics of the
production processes.
Particularly, for non-interacting identical particles, like photons, 
this technique is called intensity or particle interferometry.
In this case the correlations appear solely due to the effect of 
quantum statistics (QS) \cite{GGLP60,KP72}.
Similar effect was first used 
in astronomy to measure
the angular radii of stars by studying the dependence of the two-photon
coincidence rate on the distance between the detectors
(HBT effect \cite{hbt}).
In particle physics the QS interference was first
observed as an enhanced production of the
pairs of identical pions with small opening angles 
(GGLP effect \cite{GGLP60}). Later on, similar to astronomy,
Kopylov and Podgoretsky \cite{KP72}
suggested to study the interference effect in terms of the correlation
function.\footnote
{Note
that though both the KP and HBT methods are based on 
the QS interference, they represent just orthogonal measurements \cite{KP72}.
The former, being the momentum-energy measurement, yields the
space-time picture of the source, while the latter does the opposite.
In particular, the HBT method provides the information about the angular
size of a star  
but, of course, - no information about the star radius or its
lifetime. 
}  

The effect of QS is usually considered in the limit of a low
phase-space density such that the possible multi-particle effects can be
neglected.
This approximation seems to be justified by present experimental
data which does not point to any spectacular multi-boson effects neither
in single-boson spectra nor in two-boson correlations.
These effects can however clearly manifest themselves in some rare events
({\it e.g.}, those with large pion multiplicities) or in
the eventually overpopulated regions of momentum
space; see, {\it e.g.}, \cite{pra93,ALICE}, 
references therein and Section 3.

The particle correlations are also
influenced by the effect of particle interaction 
in the final state (FSI) \cite{koo,ll1}.
Thus the effect of the Coulomb interaction dominates the correlations
of charged particles at very small relative momenta
(of the order of the inverse Bohr radius of the two-particle system), 
respectively suppressing or 
enhancing the production of particles with like or unlike charges.
Regarding the effect of the strong FSI, it is 
quite small for pions, while for nucleons it is often a dominant one
due to the very large magnitude of the s-wave singlet scattering length
of about 20 fm.

Though the FSI effect complicates the correlation analysis,
it is an important source of information allowing one to measure
the space-time characteristics of the production process even with the
help of non-identical particles \cite{ll1,BS86}. 
Moreover, the unlike particle correlations, 
contrary to those of identical particles, 
are sensitive to the relative space-time asymmetries in their production,
{\it e.g.} - to the relative time delays,
thus giving an important information not accessible
in the standard interferometry measurements \cite{LLEN95}.
In Section 4 we briefly formulate the theory of these correlations
and demonstrate the possibilities of the corresponding correlation technique.

\section{Formalism}\label{sec:Formalism}
As usual, we will assume
sufficiently small phase-space
density of the produced multi-particle system,
such that the correlation of two particles
emitted with a small relative
velocity in nearby space-time points is influenced by the effects of
their mutual QS and FSI only.\footnote
{
This assumption may be not valid in the case of low energy heavy ion reactions
when the particles are produced in a strong Coulomb field of residual
nuclei. To deal with this field a quantum $adiabatic$
approach can be used \cite{LLEN95}.
} 
We define the ideal two-particle
correlation function $R(p_{1},p_{2})$ as a
ratio of the differential two-particle production cross section to the
reference one which would be observed in the absence of the
effects of QS and FSI. 
In heavy ion  or high energy hadronic collisions 
we can neglect kinematic constraints
and most of the dynamical correlations and construct the
reference distribution by mixing the particles from different
events. 

Assuming the momentum dependence of the one-particle emission probabilities
in- essential when varying the particle 4-momenta $p_{1}$ and $p_{2}$ by the
amount characteristic for the correlation due to QS and FSI
({\it smoothness assumption}),
{\it i.e.} assuming that
the components of the mean space-time distance
between particle emitters 
are much larger than those of the space-time extent
of the emitters, 
we get the well-known result of Kopylov and Podgoretsky for
identical particles,
modified by the substitution of the plane wave
$e^{ip_{1}x_{1}+ip_{2}x_{2}}$ by the nonsymmetrized Bethe-Salpeter
amplitudes in the continuous spectrum of the two-particle states
$\psi_{p_{1}p_{2}}^{S(+)}(x_{1},x_{2})$,
where $x_{i}=\{t_{i},{\bf r}_{i}\}$ are the 4-coordinates 
of the emission points
of the two particles and $S$ is their total spin \cite{ll1}.
At equal emission times in the
two-particle c.m.s. ($t^{*}=t_{1}^{*}-t_{2}^{*}=0$) this amplitude
coincides (up to an unimportant phase factor due to the c.m.s. motion)
with a stationary solution  of the scattering problem
$\psi_{-{\bf k}^{*}}^{S(+)}({\bf r}^{*})$,
where 
${\bf k}^{*}= {\bf p}^{*}_{1} = -{\bf p}^{*}_{2}$
and ${\bf r}^{*}= {\bf r}^{*}_{1} -{\bf r}^{*}_{2}$
(the minus sign of the vector ${\bf k}^{*}$ corresponds to the reverse
in time direction of the emission process). 
The Bethe-Salpeter amplitude can be usually substituted  by
this solution ({\it equal time} approximation).\footnote
{The {\it equal time} approximation is valid
on condition \cite{ll1}
$ |t^*|\ll m_{2,1}r^{*2}$ for 
${\rm sign}(t^*)=\pm 1$ respectively.
This condition is usually satisfied
for heavy particles like kaons or
nucleons. But even for pions, the $t^{*}=0$ approximation
merely leads to a slight overestimation (typically $<5\%$) of the strong
FSI effect and, 
it doesn't influence the leading zero--distance 
($r^{*}\ll |a|$) effect of the Coulomb FSI.}
Then, for nonidentical particles,
\begin{equation}
R(p_{1},p_{2})=
\sum_{S}\rho_{S}
\langle |\psi_{-{\bf k}^{*}}^{S(+)}({\bf r}^{*})|^{2}
\rangle _{S}.
\label{1}
\end{equation}
Here the averaging is done over the emission points
of the two particles in a state with total spin $S$
populated with the probability
$\rho_{S}$,
$\sum_{S}\rho_{S} = 1$.\footnote
{
For unpolarized particles
with spins $s_{1}$ and $s_{2}$ the probability 
$\rho_{S}=(2S+1)/[(2s_{1}+1)(2s_{2}+1)]$.
Generally, the correlation function is sensitive to particle
polarization. For example, if two spin-1/2 particles are emitted with 
polarizations ${\bf P}_1$ and ${\bf P}_2$ then \cite{ll1}
$\rho_0=(1-{\bf P}_1\cdot{\bf P}_2)/4$ and
$\rho_1=(3+{\bf P}_1\cdot{\bf P}_2)/4$.
}
For identical particles, the amplitude in Eq.~(\ref{1})
should be properly symmetrized:
\begin{equation}
\label{3}
\psi_{-{\bf k}^{*}}^{S(+)}({\bf r}^{*}) \rightarrow
[\psi_{-{\bf k}^{*}}^{S(+)}({\bf r}^{*})+(-1)^{S}
\psi_{{\bf k}^{*}}^{S(+)}({\bf r}^{*})]/\sqrt{2}.
\end{equation}
 
Particularly, for non-interacting identical particles
the characteristic feature of the correlation function is  the
presence of the interference maximum or minimum at small
$|{\bf q}|$ changing to a horizontal plateau at sufficiently
large $|{\bf q}|$. For example, assuming that particles are produced
independently
and that the space and time limitation of the  production
process is effectively described by the
one-particle probability of a Gaussian form with the corresponding
dispersions $r_0^2$ and $\tau_0^2$,
Eqs. (\ref{1}) and (\ref{3}) yield:
\begin{equation}
R(p_{1},p_{2})= 1 + \sum_{S}(-1)^{S}\rho_{S}
\exp [-r_{0}^{2}q_{T}{}^{2}-(r_{0}{}^{2}+
v^2\tau_{0}{}^2)q_{L}{}^{2}].
\label{5}
\end{equation}
Here $q_T$ and $q_L$ are the transverse and
longitudinal components of the relative 3-momentum ${\bf q}$ 
with respect to the pair velocity vector ${\bf v}$
($q_{T}=2k_T^*$ and $q_L=q_0/v=2\gamma k_L^*$, where $\gamma$ is the pair 
Lorentz factor).
We see that, due to the relation $q_0 = {\bf v}{\bf q} \equiv vq_L$,
the correlation function at $v\tau_{0} > r_{0}$ substantially depends
on the direction of the vector  ${\bf q}$
even in the case of
spherically symmetric spatial form of the production region,
the interferometric radii squared in the $T-$ and $L-$directions being
$r_T^2=r_0^2$ and $r_L^2=r_0^2+v^2\tau_0^2$.
Generally, the directional dependence of the correlation
function can be used to determine both
the characteristic emission time and the form
of the production region \cite{KP72}.

It should be noted that particle correlations at high energies
usually measure only a small part of the space-time emission volume since,
due to substantially limited decay momenta 
of few hundred MeV/c, the sources,
despite their fast longitudinal motion, emit
the correlated particles with nearby velocities mainly
at nearby points in the c.m.s. of particle pair.
The dynamical examples are sources-resonances,
colour strings or hydrodynamical expansion.
To get rid of a fast longitudinal motion, sometimes the longitudinally
comoving system (LCMS) is introduced, in which each pair is emitted
transverse to the reaction axis.

\section{Multi-boson effects}
The two-body formalism of Section 2 assumes that the phase-space
density $f$ of the produced multi-particle system is small compared to
unity so
that the multi-particle correlations can be neglected.
The mean phase-space density of the pions of a given type at a given 
momentum {\bf p} (rapidity $y$ and transverse momentum 
${\bf p}_t$) can be estimated, in the low density limit, 
as a number of the pions interfering with  
a pion of momentum {\bf p} and building the Bose-Einstein (BE)
enhancement \cite{ber94}.
Using the usual Gaussian parametrization for the correlation function in 
the LCMS:                                     
$R(p_1,p_2)=1+\lambda \exp(-r_x^2q_x^2-r_y^2q_y^2-r_z^2q_z^2)$,
and parametrizing
the cut-off of transverse momenta by an exponential law in transverse
mass $m_t$,
we can estimate the mean pion phase-space density as
\begin{equation}
\langle f\rangle _{{\bf p}}
\approx \lambda\pi^{3/2}\frac{\mathop{\rm cosh}\nolimits y}{V}
\frac{d^3n}{d^3{\bf p}}
=\lambda\frac{\sqrt\pi}{2}
\frac{\exp[-(m_t-m)/T]}{VT(T+m)m_{\perp}}\frac{dn}{dy},
\label{8.4}
\end{equation}
where $V=r_xr_yr_z$ is the interference volume. 
The mean density is maximal for soft pions ($p_t\sim y\sim 0$).
At present energies it is typically less than $\sim 0.2$ thus
indicating rather small multi-boson effects.
These effects can be expected small also in future heavy ion exeperiments
at RHIC and CERN since presently
the LCMS interference volume $V$ seems to
scale with the density $dn/dy$ 
pointing to the freeze-out of the particles
at a constant phase-space density. 
Nevertheless, the multi-boson effects can show up in certain classes
of events or in some regions of momentum space.

The multi-boson effects can be practically treated provided that we can
neglect particle interaction in the final state and assume independent
particle emission, 
supplemented by the requirement of a universal single-particle
emission function, independent of the origin of single-particle
sources. The multi-particle emission function then being 
a product of the single-particle ones. The latter allows to calculate
the Bose-Einstein effect on the original multiplicity distribution
or boson spectra with the help of the BE weights $\omega_n$
\cite{pra93}
which can be expressed through the so called cumulants  
or differential cumulants \cite{ame-led}.
The calculation of all the cumulants is generally a difficult task. 
Thus, for realistic transport models used to predict particle production in
ultra-relativistic heavy-ion collisions, the numerical 
limitations allow to determine only a few lowest order cumulants
(up to about the fourth order) \cite{ame-led}.
Fortunately, they are sufficient for typical rather moderate
pion freeze-out phase-space
densities \cite{mb98}.

The analytical calculation of the cumulants of all orders is possible
in a simple model in which the pions are instantaneously emitted according to 
the Gaussian emission function (in Wigner-like phase-space) characterized by
the dispersions $\Delta^2$ and $r_0^2$ in the momentum
and ordinary space respectively \cite{pra93}.\footnote
{
Note that in this model the two-pion correlation function
in the low density limit is given by Eq. (\ref{5}) with $S=0$,
$\tau_0=0$
and $r_0^2$ substituted by 
the dispersion $r_0^2-1/(4\Delta^2)$
of the centers of the elementary
emitters, where $1/(4\Delta^2)$ represents the minimal possible dispersion
of the Wigner-like spatial coordinates due to emitter finite sizes.}

Assuming a Poissonian distribution of the original boson multiplicities
with the mean multiplicity $\eta$,
the resulting one, at large $n$, tends to the BE distribution with the
mean multiplicity $\xi/(1-\xi)$, where $\xi=\eta/(r_0\Delta+1/2)^3$
can be considered as an intrinsic phase space parameter.
Its maximal value of 1 corresponds to the explosion of the
multiplicity distribution.
Comparing the model phase-space density 
$\langle \tilde{f} \rangle_{{\bf p}=0} \approx \eta/(\sqrt{2}r_0\Delta)^3$
with the experimental estimate of about 0.2, we get for the density 
parameter $\xi\approx 0.4-0.5$.
It was shown \cite{ALICE} that at such densities an approximate
$\xi$-scaling of the ratio $\langle n\rangle/\eta$ takes place;
$\langle n\rangle$ becomes close to the asymptotic scaling value of
$\xi/(1-\xi)$ only at  $\xi>0.9$.

Regarding the influence of the BE effect on the single-boson
spectrum, at sufficiently large momenta, when the
local density 
$\langle f\rangle_{{\bf p}}$
remains small even at large multiplicities, 
it is dominated by
the contribution 
of the original spectrum, 
otherwise, at large local densities, it is determined by the asymptotic
large-density spectrum
with the original dispersion $\Delta^2$ substituted by $\Delta/(2r_0)$
\cite{ALICE}.
Experimentally the effect of BE "condensate" 
was searched for at SPS CERN as a low-$p_t$
enhancement, however, with rather uncertain results.

\begin{figure}
\epsfig{file=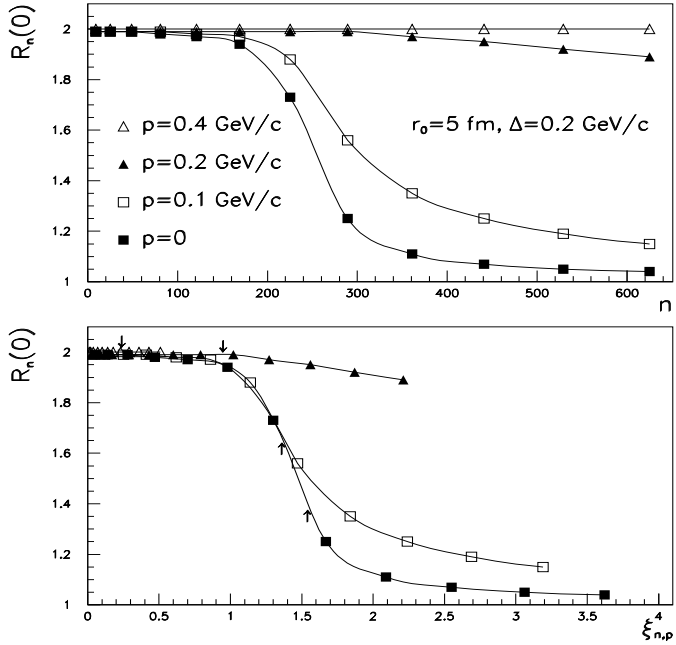, width=10.cm, height=10.cm}
\caption{ The intercept of the two-pion correlation functions
as a function of the multiplicity $n$
and the density
parameter $\xi_{n,{\bf p}}$ for several values 
$p=0$, 0.1, 0.2 and 0.4 GeV/c 
of the mean momentum of the two pions.
The arrows on the interpolating curves 
indicate the intercept values corresponding to 
$\xi_n=3\xi=1.5$ ($n\approx 3\langle n\rangle$).}
\label{intercept}
\end{figure} 

As for the two-boson correlation function, 
it is well known that
for a given multiplicity $n$, the intercept $R_n(0)$
decreases and the correlation function width increases
with the increasing $n$ 
or decreasing momentum $p$,
both corresponding to the increasing local density.
In Fig. 1 we show \cite{mb98} 
the intercept as a function of the multiplicity $n$
and the density
parameter $\xi_{n,{\bf p}}=\xi_n\exp(-{\bf p}^2/2\Delta^2)$,
$\xi_n=n/(r_0\Delta+1/2)^3$,
for several values of the mean momentum:
$p=0$, 0.1, 0.2 and 0.4 GeV/c. 
As expected, the intercept is practically constant at low local
densities (small $n$ or high $p$). As condensate develops, the intercept
sharply falls down. The sharpness of this drop is however less
pronounced at higher momenta even if plotted as a function of the
local density.
Clearly, this lack of density scaling is related to
a strong decrease of the condensate contribution with the increasing
momentum.
To demonstrate the possibility of the observation 
of the condensate effect for soft pions produced in certain high multiplicity
events not following the ordinary proportionality rule between
the freeze-out phase space density and pion multiplicity, 
in Fig. 1 we indicate by the arrows 
the intercept values corresponding to 
$\xi_n=3\xi=1.5$ ($n\approx 3\langle n\rangle$).

In the inclusive case corresponding to the original Poissonian
multiplicity distribution,
the correlation function intercept is equal to 2 for any pion
momenta or the local densities \cite{ALICE} and its width
logarithmically increases with the increasing condensate
contribution \cite{mb98}.

The results of the considered model should not be taken, however,
too literally since: 

a) in contradiction with the experimental indications on 
a constant freeze-out phase space density, 
in the model there is no correlation 
between the emission volume and pion multiplicity;

b) the static character of the model can be justified
(neglecting the transverse expansion)
in a limited rapidity region only. Thus the pions with a rapidity 
difference greater than about unity have to be considered
as originating from different static sources;

c) since the inclusive spectra at sufficiently high mean multiplicity
are dominated by the condensate contributions 
in both extreme cases of very narrow or very wide multiplicity
distributions of the originally
uncorrelated bosons, the intercept of the inclusive
correlation function is likely to be less than the value of 2
corresponding to the Poissonian case;

f) for identical charged pions, the BE effects are suppressed due to the
Coulomb repulsion. Since this repulsion is important only
in a weakly populated region of very small relative momenta,
the suppression of the global BE weights $\omega_n$ is rather small.
For example, for $\omega_2$ this suppression, being about
$(a r_0\Delta^2)^{-1}$, is usually less than one per mill.
The Coulomb distortion of the global multi-boson effects is
therefore negligible in the rare gas limit.
Nevertheless, since
the Coulomb repulsion destroys the formation of the condensates
made up from positive and negative pions in the disjoint
phase space regions, it can lead to noticeable differences
between charged and neutral pions
in dense systems.
Particularly, we can expect a decrease of the
charge-to-neutral multiplicity ratio with the increasing
phase space density.

\section{Measuring the relative space-time asymmetries}

The correlation function of two nonidentical particles, compared with the
identical ones,
contains a principally new piece of information on the relative
space-time asymmetries in particle emission \cite{LLEN95}.
This is clearly seen in the case of neutral particles
when the two-particle amplitude
$\psi_{-{\bf k}^{*}}^{S(+)}({\bf r}^{*})$
takes on the form
\begin{equation}
\label{6}
\psi_{-{\bf k}^{*}}^{S(+)}({\bf r}^{*}) =
e^{-i{\bf k}^{*}{\bf r}^{*}}+\phi^{S}_{k^{*}}(r^{*}),
\end{equation}
where the scattered wave $\phi^{S}_{k^{*}}(r^{*})$, in the considered
region of small relative momenta, is independent of the
directions of the vectors ${\bf k}^{*}$ and ${\bf r}^{*}$.
Inserting Eq. (\ref{6}) into the formula (\ref{1}) for 
the correlation function, we can see that the latter 
is sensitive to the relative space-time asymmetry due to the
odd term $\sim\sin {\bf k}^{*}{\bf r}^{*}$. 
Particularly, it allows for a measurement of the mean relative delays
$\langle t\rangle\equiv\langle t_1-t_2\rangle$ in particle emission.
To see this, let us make the Lorentz
transformation from the rest frame of the source
to the c.m.s. of the
two particles:
$r_{L}^{*} = \gamma (r_{L}-vt)$, $r_{T}^{*}=r_{T}$.
Considering, for simplicity, the behavior of the vector ${\bf r}^{*}$
in the limit $|vt| \gg r$, 
we see that this vector
is only slightly affected by averaging over the spatial
distance $r \ll |vt|$ of the emission points in the rest frame
of the source so that 
${\bf r}^{*}\approx -\gamma{\bf v}t$.
Therefore, the vector ${\bf r}^{*}$
is nearly parallel or antiparallel to the velocity vector
${\bf v}$ of the pair, depending on the sign of the time
difference $t\equiv\Delta t=t_1-t_2$.
The sensitivity to this sign is transferred to the correlation
function through the odd in
${\bf k}^{*}{\bf r}^{*}\approx  -\gamma{\bf k}^{*}{\bf v}t$ term 
provided the sign of the scalar product
${\bf k}^{*}{\bf v}$ is fixed.

For charged particles there arise additional odd terms due to
the confluent hypergeometrical function 
$F(\alpha,1,z)=1+\alpha z/1!^2+\alpha(\alpha+1)z^2/2!^2+\dots$, 
modifying the plane wave in Eq.~(\ref{6}):
\begin{equation}
\psi_{-{\bf k}^{*}}^{S(+)}({\bf r}^{*}) = {\rm e}^{i\delta}
\sqrt{A_{c}(\eta)}\left[
{\rm e}^{-i{\bf k}^{*}{\bf r}^{*}}F\left(-i\eta,1,i\rho\right)
+\phi^{S}_{ck^{*}}(r^{*})\right],
\label{14}
\end{equation}
where 
$\rho={\bf k}^*{\bf r}^*+{\rm k}^*r^*$, $\eta=(k^{*}a)^{-1}$,
$a=(\mu z_{1}z_{2}e^{2})^{-1}$ is the Bohr radius of the
two-particle system taking into account the sign of the interaction
($z_ie$ are the particle electric charges,
$\mu$ is their reduced mass),
$\delta=\mbox{arg}\Gamma(1+i\eta)$
is the Coulomb s-wave shift and
$A_{c}(\eta)=2\pi\eta/[\exp (2\pi\eta)-1]$
is the Coulomb penetration factor.\footnote
{This factor
substantially deviates from unity
only at $k^{*}< 2\pi/|a|$ ({\it e.g.}, at
$k^{*} < 22$ MeV/c for two protons).
Note that for the distances $r^*>|a|$ the 
confluent hypergeometrical function becomes important and
compensates the deviation of the Coulomb factor from unity
except for the classically
forbidden region of $k^{*} <  (|a|r^{*}/2)^{-1/2}$,
narrowing with the increasing $r^*$.
Thus the FSI practically vanishes if at least one of the two particles 
comes from a long lived
source ($\eta, \eta', \Lambda, K^0_s, \ldots$).
}
Clearly, at a given distance $r^*$, the effect of the odd component
in the Coulomb wave function is of increasing importance with
a decreasing Bohr radius of the particle pair,
{\it i.e.} for particles
of greater masses or electric charges.
At low energies,
the sensitivity of the correlation to the odd component can be somewhat
modified due to Coulomb interaction with the residual charge \cite{LLEN95}.
 
It is clear that in the case of a dominant time asymmetry, 
$v|\langle t\rangle| \gg |\langle r_L\rangle|$,
a straightforward way to determine
the mean time delay $\langle t\rangle $ is to
measure the correlation functions
$R_{+}({\bf k}^{*}{\bf v}\geq 0)$ and
$R_{-}({\bf k}^{*}{\bf v}< 0)$.
Depending on the sign of $\langle t\rangle$,
their ratio $R_{+}/R_{-}$ should show a peak
or a dip in the region of small $k^{*}$ and approach 1
at large values of $k^{*}$.
As the sign of the scalar product
${\bf k}^{*}{\bf v}$ is practically equal to that of the
difference of particle velocities $v_{1}-v_{2}$
(this equality is always valid for particles of equal masses),
the sensitivity of the correlation functions
$R_{+}$ and $R_{-}$ to the sign of the difference of particle
emission times has a simple classical explanation. Clearly,
the interaction between the particles in the case of an earlier
emission of the faster particle will be weaker compared
with the case of its later emission (the interaction time being
longer in the latter case leading to a stronger correlation).
This expectation is in accordance with 
Eqs. (\ref{1}) and (\ref{14})
at $k^* \rightarrow 0$, $\langle r^*\rangle \ll |a|$ and
$\langle |\phi^{S}_{ck^{*}}(r^{*})|\rangle \ll 1$, when
(the arrow indicates the limit
$v|\langle t\rangle| \gg |\langle r_L\rangle|$):
\begin{equation}
R_+/R_-\approx 1+
2\frac{\langle r_L^*\rangle }{a} \rightarrow
1-2\frac{\langle \gamma v(t_1-t_2)\rangle }{a}.
\label{51}
\end{equation}
 
The sensitivity of the $R_+/R_-$ correlation method to the mean
relative time shifts (introduced {\it ad hoc}) 
was studied \cite{ALICE}
for various two-particle systems 
simulated in $Pb+Pb$ collisions at SPS energy
using the event generator VENUS 5.14 \cite{WER93}.
The scaling of the effect with the space-time asymmetry and 
with the inverse Bohr radius $a$,
indicated by Eq. (\ref{51}), was clearly illustrated
for the $K^+K^-$ system ($a=-110$ fm)
and for the like- and unlike-sign $\pi K$, $\pi p$ and $Kp$
systems ($a=\pm 249$, $\pm 226$ and $\pm 84$ fm respectively).                                                                  
It was concluded that for sufficiently relativistic pairs
($\gamma v > 0.5$)
the $R_{+}/R_{-}$ ratio can be sensitive to the shifts in the
particle emission times of the order of a few fm/c. 
Motivated by this result the $R_+/R_-$ method was  recently 
applied to the $K^+K^-$ system
simulated in a two-phase thermodynamical evolution model
and the sensitivity was demonstrated
to the production of the transient strange quark matter state
even if it decays on strong interaction time scales \cite{sof97}.

\begin{figure}[hb] 
\begin{center}\mbox{ 
                \epsfxsize=9cm
                 \epsffile{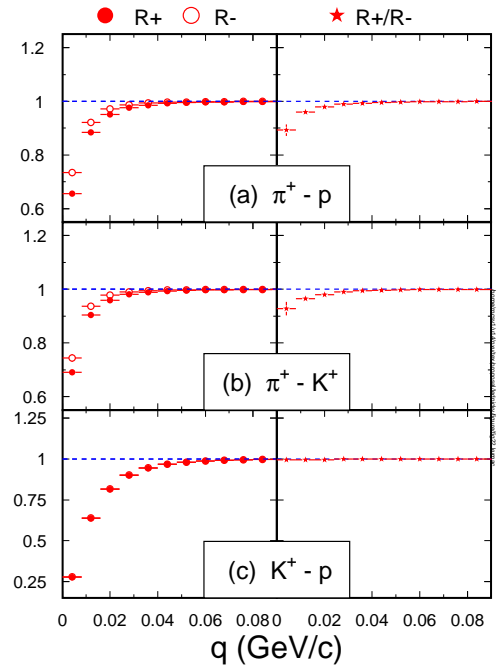}
}
\end{center}
\caption{ 
Unlike particle correlation functions 
$R_+$ and $R_-$ and their ratios simulated 
with RQMD
for mid-rapidity particle
pairs $\pi^+p$, $\pi^+K^+$, and $K^+p$ 
in $Pb+Pb$ collisions at SPS energy.
}
\label{pikpSPS}
\end{figure}

The method sensitivity was also studied 
for AGS and SPS energies using the transport code RQMD v2.3 \cite{rqmd}.
Thus at SPS energy the central $Pb+Pb$ collisions have been simulated
and the $\pi^+K^+$, $\pi^+p$ and $K^+p$ correlations 
have been studied \cite{lpx}. 
To get rid of the effect of a fast longitudinal motion,
the study was done in the longitudinally co-moving system (LCMS)
in which the pair is emitted transverse
to the reaction axis so that $v=v_{\perp}$, $r_L=\Delta x$ and
\begin{equation}
\label{delta}
\Delta x^*=\gamma_{\perp}(\Delta x-v_{\perp}\Delta t),~~\Delta y^*=\Delta y,~~
\Delta z^*=\Delta z.
\end{equation}
The simulated correlation functions $R_+$, $R_-$ and their ratios are plotted 
in Fig. \ref{pikpSPS}.
We can see that for $\pi ^+p$ and $\pi ^+K^+$ systems these ratios are less than
unity at small values of $q\equiv k^*$, 
while for $K^+p$ system the ratio $R_+/R_-$
practically coincides with unity. These results well agree with the mean values
of $\Delta t$,  $\Delta x$ and $\Delta x^*$ presented in Table 1
($\langle\Delta y\rangle\approx\langle\Delta z\rangle\approx 0$).
It can be seen from Eqs. (\ref{51}) and (\ref{delta}) 
that the absence of the effect in the $R_+/R_-$ ratio for the  $K^+p$
system is due to practically the complete compensation of the space and time
asymmetries leading to $\Delta x^* \approx 0$. For $\pi^+p$ system the effect
is determined mainly by the x-asymmetry. For $\pi^+K^+$ system both the
x- and time-asymmetries contribute in the same direction, 
the latter contribution
being somewhat larger.
The separation of the relative time delays from the
spatial asymmetry is, in principle, possible (see Eq. (\ref{delta}))
by studying the ratio $R_+/R_-$ in different intervals of the
pair velocity.

At AGS energy 
the $Au+Au$ collisions have been simulated and 
the $\pi^+p$ correlations have been studied
in the projectile fragmentation region where proton directed
flow is most pronounced and where the proton and pion sources
are expected to be shifted \linebreak
\newpage
\begin{table}[htbp]
\caption[]{\footnotesize
Mean values of the relative space-time coordinates in LCMS
(in fm) calculated \cite{lpx} from
RQMD (v2.3) for 158 A$\cdot$GeV $Pb+Pb$ central collisions.
}
\label{trq}

\medskip
\begin{center}
\begin{tabular}{|c|c|c|c|c|c|} 
\hline
     &system & $\langle\Delta t\rangle$ & $\langle\Delta x\rangle$
     & $\langle\Delta x -v_{\perp}\Delta t\rangle$
     & $\langle\Delta x^*\rangle$\\
\hline
        &$\pi^+p$   &-0.5 &-6.2 &-6.4 &-7.9\\
        &$\pi^+K^+$ & 4.8 &-2.7 &-5.8 &-7.9\\  
        &$K^+p$    &-5.5 &-3.2 &-0.6 &-0.4\\
\hline
\end{tabular}
\end{center}
\end{table}

\newpage
\noindent
relative to each other both in the longitudinal
and in the transverse (flow) directions
in the reaction plane.
It was demonstrated that a modification
of the $R_{+}/R_{-}$ method ($\pm$ corresponding now to the signs of the
respective components $k^*_i$) is sufficiently sensitive to reveal these
shifts \cite{vol97}.
Recently, the shifts predicted by RQMD were confirmed by applying the
$R_{+}/R_{-}$ method to the experimental AGS data \cite{mis98}.

At low energies, the particles in heavy ion collisions are emitted
with the characteristic emission times of tens to hundreds fm/c so that
the observable time shifts should be of the same order \cite{LLEN95}.
In fact the $R_{+}/R_{-}$ method has been successfully applied
to study proton-deuteron correlations in several heavy ion
experiments at GANIL \cite{ghi95,nou96}. 
It was observed, in agreement 
with the coalescence model, that deuterons are on average emitted earlier
than protons.

\section{Conclusion}
Using the analytically solvable Gaussian model,
we have discussed
the influence of the multi-boson effects on boson multiplicities,
single-boson spectra and two-boson correlations,
including an approximate scaling behavior 
of some of their characteristics with the
phase space density.
Though these effects are hardly to be observable in typical events
of heavy ion collisions in present and perhaps also in future 
heavy ion experiments, they can show up in certain classes
of events,
{\it e.g.}, in those with high pion multiplicities.

We have shown that unlike particle
correlations, compared with those of
identical particles,
contain a principally new piece of information on the relative
space-time asymmetries in particle emission, thus allowing,
in particular, a measurement of the mean relative delays
in particle emission at time scales as small as $10^{-23}$ s. 
To determine these asymmetries, the unlike particle
correlation functions $R_+$ and $R_-$ have to be studied
separately for positive and negative values of the projection
of the relative momentum vector in pair c.m.s. on the 
pair velocity vector or, generally, - on any direction of interest.
We have discussed here the results of recent studies of these
correlation functions for
a number of two-particle systems simulated with various event generators.
It was shown
that the $R_+/R_-$ ratio is sufficiently sensitive to the
relative space-time asymmetries arising
due to the formation of the quark-gluon plasma and strangeness
distillation and even to those expected in the usual dynamical scenarios
at AGS and SPS energies. 
As to the detection of the unlike particles with close velocities 
($p_1/m_1\approx p_2/m_2$),
there is no problem with the two-track resolution since these
particles,
having either different momenta or different charge-to mass ratios,
have well separated trajectories in the detector magnetic field.
For the same reason, however, a large momentum acceptance of the
detector is required.

This work was supported by GA AV Czech Republic, Grant No. A1010601
and by GA Czech Republic, Grant No. 202/98/1283.

%

\newpage
\begin{thebibliography}{99}
\bibitem{GGLP60}
{G.~Goldhaber et al.:} {\it Phys. Rev.} {\bf 120} (1960) {300};
\bibitem{KP72}
{G.I.~Kopylov, M.I.~Podgoretsky:} {\it Yad.~Fiz.} {\bf 15} (1972) {392}
({\it Sov. J. Nucl. Phys.} {\bf 15} (1972));
\refer{G.I.~Kopylov}{Phys. Lett.}{50}{1974}{472}
{M.I.~Podgoretsky:}
{\it Fiz.~Elem. Chast. Atom. Yad.} {\bf 20} (1989) {628}
({\it Sov. J. Part. Nucl.} {\bf 20} (1989) {266});

\bibitem{hbt}
R.~Hanbury-Brown, R.Q.~Twiss:
{\it Phil. Mag.} {\bf 45} (1954) 663;
{\it Nature} {\bf 178} (1956) 1046;

\bibitem{pra93}
\refer{S.~Pratt}{Phys. Lett. B}{301}{1993}{159}
{\it Phys. Rev. C} {\bf 50} (1994) {469};

\bibitem{ALICE} B. Erazmus et al.: Internal Note ALICE 95-43, Geneva 1995;

\bibitem{koo}
\refer{S.E.~Koonin}{Phys. Lett. B}{70}{1977}{43} 
\refer{M.~Gyulassy, S.K.~Kauffmann, L.W.~Wilson}
{Phys. Rev. C}{20}{1979}{2267}

\bibitem{ll1}
R.~Lednicky, V.L.~Lyuboshitz:
{\it Yad.~Fiz.} {\bf 35} (1982) 1316 ({\it Sov. J. Nucl. Phys.} 
{\bf 35} (1982) 770);
Proc. CORINNE 90,
Nantes, France, 1990 (ed. D.~Ardouin, World Scientific, 1990) p. 42;
JINR report P2-546-92 (1992); 
{\it Heavy Ion Physics} {\bf 3} (1996) 93;

\bibitem{BS86}
\refer{D.H.~Boal, J.C.~Shillcock} 
{Phys. Rev. C}{33}{1986}{549}
\refer{D.H.~Boal, C.-K. Gelbke, B.K.~Jennings}
{Rev. Mod. Phys.}{62}{1990}{553}

\bibitem{LLEN95}
\refer{R.~Lednicky, V.L.~Lyuboshitz, B.~Erazmus, D.~Nouais}
{Phys. Lett. B}{373}{1996}{30}
Report SUBATECH 94-22, Nantes 1994;

\bibitem{ber94}
G.F.~Bertsch,
{\it Phys.~Rev.~Lett.} {\bf 72} (1994) 2349.

\bibitem{ame-led}
N.S.~Amelin, R.~Lednicky:
{\it Heavy Ion Physics} {\bf 4} (1996) 241;
SUBATECH 95-08, Nantes 1995;

\bibitem{mb98}
R.~Lednicky et al.: Multi-boson effects,
to be published;

\bibitem{WER93} \refer{K.~Werner}
{Phys.~Rep.}{232}{1993}{87}  
\refer{K. Werner and J. Aichelin}
{Phys. Rev. C}{52}{1995}{1584} 

\bibitem{sof97} S.~Soff et al.:
{\it J.~Phys. G} {\bf 12} (1997) 2095;

\bibitem{rqmd} \refer{H.~Sorge et al.}
{Phys. Lett. B}{243}{1990}{7};

\bibitem{lpx} R.~Lednicky, S.~Panitkin, Nu Xu:
submitted to QM97, Tsukuba;

\bibitem{vol97} \refer{S.~Voloshin, R.~Lednicky, S.~Panitkin, Nu Xu}
{Phys. Rev. Lett.}{79}{1997}{4766}

\bibitem{mis98} D.~Miskowiec: nucl-ex/9808003;

\bibitem{ghi95} \refer{C.~Ghisalberti et al.}
{Nucl. Phys. A}{583}{1995}{401}

\bibitem{nou96} B. Erazmus et al.: Proc. XXXIV Bormio Meeting (1996);
D.~Nouais, PhD Thesis (1996), Nantes;
D.~Gourio, PhD Thesis (1996), Nantes.
%
%
\end{thebibliography}
\end{document}